\documentclass[conference]{IEEEtran}
\IEEEoverridecommandlockouts
\usepackage{cite}
\usepackage{amsmath,amssymb,amsfonts}
\usepackage{algorithmic}
\usepackage{graphicx}
\usepackage{textcomp}
\usepackage{hyperref}
\usepackage{scrextend}
\usepackage{soul}
\usepackage{float}
\usepackage{caption}
\usepackage{subcaption}

\usepackage{xcolor}
\def\BibTeX{{\rm B\kern-.05em{\sc i\kern-.025em b}\kern-.08em
    T\kern-.1667em\lower.7ex\hbox{E}\kern-.125emX}}
\begin{document}

\title{Incremental Attractor Neural Network Modelling of the Lifespan Retrieval Curve
}

\author{\IEEEauthorblockN{Patrícia Pereira}
\IEEEauthorblockA{
\textit{INESC-ID}\\
\textit{Instituto Superior Técnico}\\
\textit{Universidade de Lisboa}\\
Lisbon, Portugal \\
patriciaspereira@tecnico.ulisboa.pt}
\and
\IEEEauthorblockN{Anders Lansner}
\IEEEauthorblockA{
\textit{Computational Brain Science Lab}\\
\textit{KTH Royal Institute of Technology}\\
\textit{Stockholm University}\\
Stockholm, Sweden \\
ala@kth.se}
\and
\IEEEauthorblockN{Pawel Herman}
\IEEEauthorblockA{
\textit{Computational Brain Science Lab}\\
\textit{KTH Royal Institute of Technology}\\
Stockholm, Sweden \\
paherman@kth.se}

}

\maketitle

\begin{abstract}
The human lifespan retrieval curve describes the proportion of recalled memories from each year of life. It exhibits a reminiscence bump - a tendency for aged people to better recall memories formed during their young adulthood than from other periods of life. We have modelled this using an attractor Bayesian Confidence Propagation Neural Network (BCPNN) with incremental learning. We systematically studied the synaptic mechanisms underlying the reminiscence bump in this network model after introduction of an exponential decay of the synaptic learning rate and examined its sensitivity to network size and other relevant modelling mechanisms. The most influential parameters turned out to be the synaptic learning rate at birth and the time constant of its exponential decay with age, which set the bump position in the lifespan retrieval curve. The other parameters mainly influenced the general magnitude of this curve. Furthermore, we introduced the parametrization of the recency phenomenon - the tendency to better remember the most recent memories - reflected in the curve’s upwards tail in the later years of the lifespan. Such recency was achieved by adding a constant baseline component to the exponentially decaying synaptic learning rate. 
\end{abstract}


\section{Introduction}

Memory plays a key role within cognitive neuroscience and psychology. Its study has gained increased importance with the growing interest in tackling neurological and psychiatric diseases of which memory deficits are common symptoms. Memory is also a key aspect of cognition fundamental for intelligent behavior, for instance, in perception, reasoning, and decision processes. It is therefore also a crucial component of artificial intelligence systems with more advanced capabilities.

Our study concerns the modelling of long-term memory, more precisely episodic memory. It is established through long-term potentiation and depression, by which synaptic connections between neurons in the brain are strengthened or weakened thereby forming memory specific cell assemblies \cite{malenka2004ltp}. There are several memory systems acting on different time scales, where long-term memory is important because it concerns the ability to learn new information and to recall that information later in time.


\begin{figure}[H]
  \includegraphics[width=0.99\linewidth]{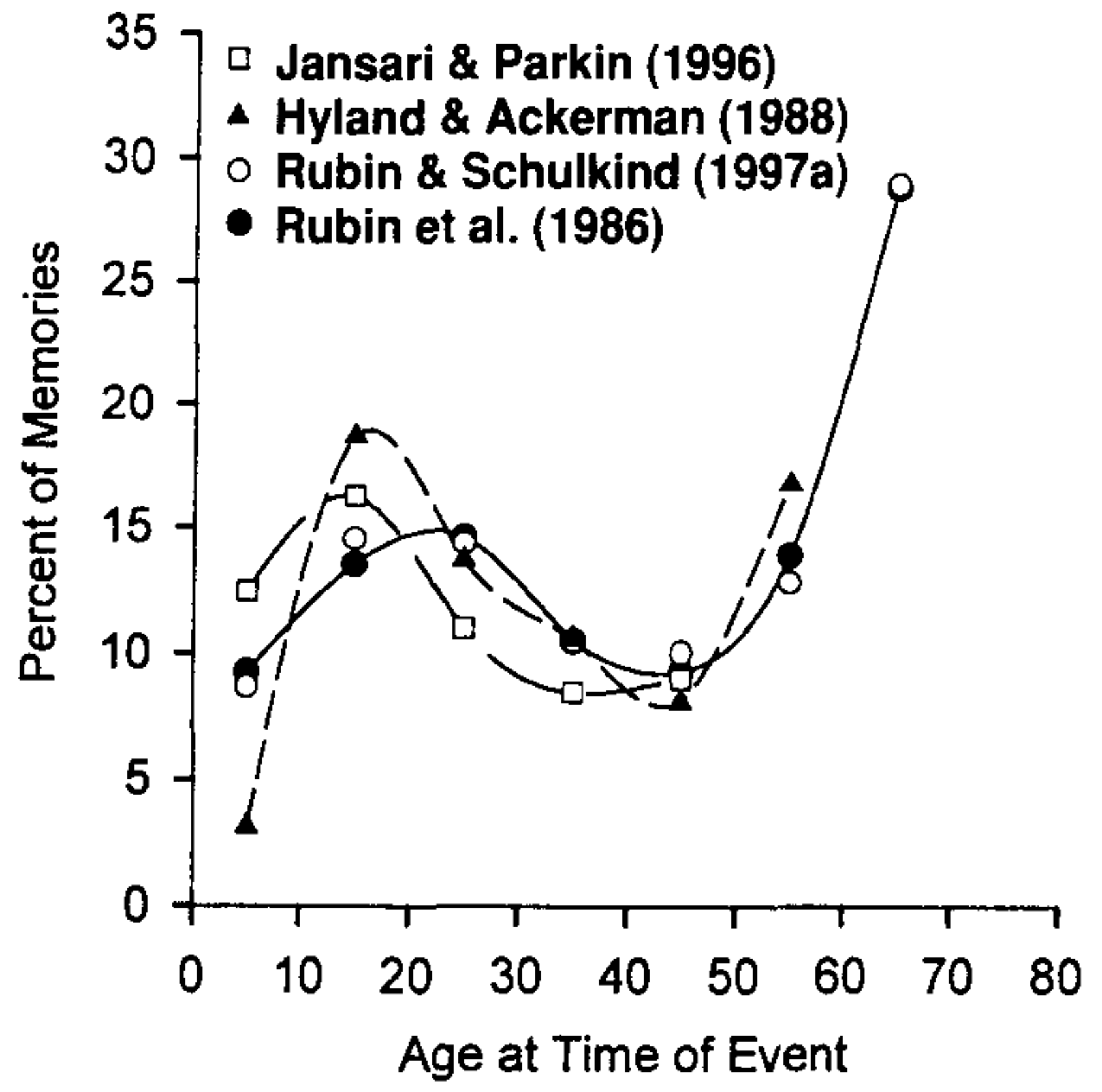}
  \caption{The lifespan retrieval curve  \cite{rubin1998things}}
  \label{f1}
\end{figure}

We focus here on the encoding of episodic memories over the course of a human lifetime that can be described by
the lifespan retrieval curve obtained in experimental studies, depicted in Figure \ref{f1}.

The curve exhibits a reminiscence bump - a tendency for aged people to recall more memories formed during their their 15-30 years of age than from other periods of life - that has consistently been observed in autobiographical memory research \cite{5}. It also displays childhood amnesia - the inability of adults to recall memories from their childhood \cite{alberini2017infantile}.

We use the firing-rate based Bayesian Confidence Propagation Neural Network (BCPNN) with incremental learning \cite{39} to model this phenomenon, by modulating its learning rate as an exponentially decaying plasticity function \cite{8}. 

Feed-forward BCPNNs were previously used for classification \cite{lansner1996higher} and data mining \cite{orre2000bayesian}. The attractor properties of the network were yielded by implementing it as a recurrent network \cite{39}. 

The network model’s mechanisms can be interpreted in the context of neurobiological effects. Therefore our study can provide insights helping in understanding long-term effects of episodic memory recall in the healthy brain, its decline with age and different forms of dementia.

We performed a systematic study of the mechanisms underlying the reminiscence bump in this BCPNN incremental network model. The modelled temporal evolution of synaptic plasticity, network architecture and other relevant modelling mechanisms used to replicate the reminiscence bump were systematically investigated. Specifically, we studied the influence on the size and shape of the reminiscence bump of varying each model parameter individually.

Besides the reminiscence bump, the lifespan retrieval curve exhibits another relevant property: the recency phenomenon, the tendency to remember the most recent memories, reflected in the curve's upwards tail in the later years of the lifespan. In the final part of this study, we introduced the parametrization of this recency phenomenon by adding a constant component to the exponentially decaying plasticity function in the model. This constitutes an experimentally based and biologically motivated hypothesis regarding the evolution of the brain's plasticity throughout lifespan.

\section{Related Work}

\subsection{Modelling Associative Memory}

A Hopfield net as a classical representative of attractor memory networks \cite{36} has been suggested as a model of cortical associative memory \cite{37}. It learns and stores a set of training patterns as stable attractor states, which can be retrieved even if the cue is noisy or distorted. The training algorithm relies on the Hebbian rule commonly referred to as “fire together, wire together”, i.e. the weights between neurons that are active at the same time are strengthened during training \cite{9}.

\subsection{Modelling the Lifespan Retrieval Curve}

Amongst the few works attempting to model the lifespan retrieval curve, the Memory Chain Model \cite{6} of Jansen et al. is composed of a cascade of memory stores.
When a new memory is encoded, a certain number of representations are formed in the first memory store. With time, the number of such representations declines and some are transferred to a subsequent store. Each store has its own decline rate and the stores are organized in order of decreasing decline rate, representing the consolidation of short-term memories into long-term. The strength of a memory is proportional to the number of representations it has.
Jansen et al. used this model to replicate the reminiscence bump  considering the differentiation of memory distribution into two separate functions, a decline function and an encoding-sampling function \cite{42}.

Another model targeting the lifespan retrieval curve is the Autobiographical Memory-Adaptive Resonance Theory (AM-ART) \cite{7}. AM-ART is a three-layer neural network. The event-specific knowledge is presented to the bottom layer to encode life events in the middle layer while a sequence of related events in the middle layer are encoded into an episode in a final layer. There is a flow of memory search and readout throughout all layers. This model was successfully used to model the lifespan retrieval curve originating a curve with the reminiscence bump and also childhood amnesia and the recency phenomenon.

Our approach is closely related to Hopfield Networks in the sense that we also model memory encoding, recall and forgetting with an attractor neural network. Palimpsest memory properties have also been shown for Hopfield networks with clipped weights \cite{storkey1998palimpsest}. To simulate the lifespan retrieval curve phenomena we make use of a function for the learning rate parameter that decreases exponentially with time, as does the recall probability function in the Memory Chain Model's attempt to replicate this curve. 

\section{Methods}

We begin by describing the attractor neural network model used in this study, the firing-rate BCPNN with incremental learning \cite{39}, with regards to its modularity, learning dynamics and meaning of parameters. We also present how we modelled the recency phenomenon and then proceed to explain the simulation protocol used to model the reminiscence bump along with the operationalizations involved. 

\subsection{Network Architecture}

\begin{figure}[!ht]
\begin{center}
  \includegraphics[width=0.82\linewidth]{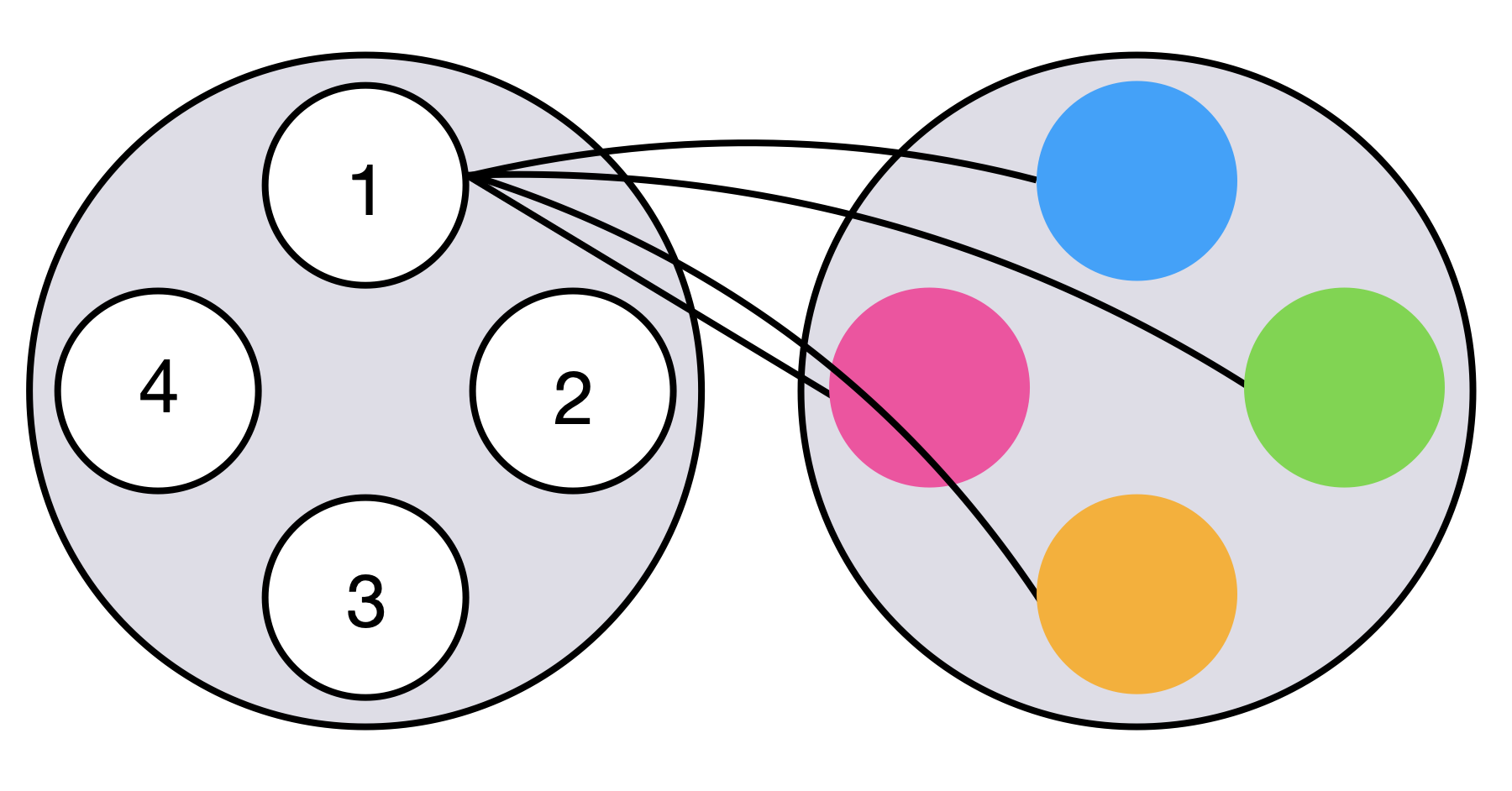}
  
 \end{center}
  \caption{Network modularity. The hypercolumn on the left (bigger circle) represents the size of a given seen object and each minicolumn (smaller circles) represents a number, meaning, for instance, \textit{small}, \textit{medium}, \textit{large} and \textit{giant}. The hypercolumn on the right represents color and each of its minicolumns represents a different color. The activity within each hypercolumn is normalized by divisive lateral inhibition. Minicolumns belonging to different hypercolumns have connections described by weights as it is depicted in the connections of the minicolumn representing the number 1.
}
  \label{f12}
  \end{figure}

The model has a specific modularity. In this network, a unit $\pi_{ii^{'}}$ (Eq. \ref{eqm1}) corresponds to activity in a minicolumn, which is a local group of neurons that is considered an elementary unit of the cortex. Minicolumns are then organized in groups, the hypercolumns. While a hypercolumn represents a discrete coded attribute or property of a memory, its corresponding minicolumns represent different values that the attribute can take as it can be seen in Figure \ref{f12} which provides an example of encoding of an object represented by two attributes.

\subsection{Network Dynamics}

The differential equations governing unit behavior in the model are:

\begin{equation}\label{eqm1}
\frac{dh_{ii^{'}}(t)}{dt}=\beta_{ii^{'}}(t)+\sum_j^Nlog\left(\sum_{j^{'}}^{M_i}w_{ii^{'}jj^{'}}(t)\pi_{jj^{'}}(t)\right)-h_{ii^{'}}(t)
\end{equation}

\begin{equation}\label{eqm2}
\pi_{ii^{'}}(t)=\frac{e^{h_{ii^{'}}}}{\sum_j e^{h_{ij}}}
\end{equation}

\begin{equation}\label{eqm3}
\frac{d\Lambda_{ii^{'}}(t)}{dt}=\alpha\left(\left[(1-\lambda_0)\hat{\pi_{ii^{'}}}(t)+\lambda_0\right]-\Lambda_{ii^{'}}(t)\right)
\end{equation}

\begin{equation}\label{eqm4}
\frac{d\Lambda_{ii^{'}jj^{'}}(t)}{dt}=\alpha\left(\left[(1-\lambda_0^2)\hat{\pi_{ii^{'}}}(t)\hat{\pi_{jj^{'}}}(t)+\lambda_0^2\right]-\Lambda_{ii^{'}jj^{'}}(t)\right)
\end{equation}

\begin{equation}\label{eqm5}
\beta_{ii^{'}}(t)=log(\Lambda_{ii^{'}}(t))
\end{equation}

\begin{equation}\label{eqm6}
w_{ii^{'}jj^{'}}(t)=\frac{\Lambda_{ii^{'}jj^{'}}(t)}{\Lambda_{ii^{'}}(t)\Lambda_{jj^{'}}(t)}
\end{equation}

A set of active units, one per hypercolumn, indexed by $i$, represents an activated memory and its level of activation, indexed by $i^{'}$ is a confidence estimate. The unit support is $h_{ii^{'}}$ and evolves according to Eq. \ref{eqm1}. 

The encoding of a memory consists in the modification of the network’s weights, $w$, and biases, $\beta$, (Eq. \ref{eqm5} and \ref{eqm6}) so that the configuration of unit activations corresponding to that memory becomes an attractor state of the network. The weights and biases (Eq. \ref{eqm6} and \ref{eqm5}, respectively) are estimated with exponential moving averages $\Lambda_{ii^{'}}$ and $\Lambda_{ii^{'}jj^{'}}$ of activity and co-activity of the connected unit activations  (Eq. \ref{eqm3} and \ref{eqm4}, respectively). The advantage of this is that the learning rule can be applied online and the network exhibits palimpsest properties which prevents catastrophic forgetting. It is therefore possible to mimic learning and gradual forgetting over time.

The BCPNN learning rule is Hebbian as the Hopfield rule, but in addition it features long-term depression (LTD), i.e. activity of the pre- or postsynaptic unit on its own leads to weakening of the weights.

While connection strengths between minicolumns belonging to different hypercolumns are represented by weights, minicolumns belonging to the same hypercolumn are related to each other via divisive lateral inhibition, implemented with softmax, in each hypercolumn (Eq. \ref{eqm2}).

There is a learning rate parameter in the incremental model, $\alpha$ (Eq. \ref{eqm3} and \ref{eqm4}), which controls the strength of encoding of each memory or how much it modifies the network’s weights and biases.

\subsection{Meaning of Model Parameters
}
\label{ssr1}

\begin{itemize}
    \item Learning rate, $\alpha$: models the degree of synaptic plasticity. By decreasing it over time during learning it can be used to represent decrease of dopamine receptor density combined with other aging phenomena, allowing the modelling of a reminiscence bump \cite{8}. This way, 
    \begin{equation}\label{alpha}
    \alpha = \alpha_0e^{-\frac{t}{\tau_s}} + \alpha_{baseline}
    \end{equation}
    with $\tau_s$  being the time constant of the age-dependent plasticity decay, that mediates these aging phenomena. In an effort to model the recency phenomenom we set this learning rate decay to stop at a certain age which yielded the desired recency tail in the retrieval curve. Regarding model equations, we parameterized this by decomposing the evolution of the learning rate as a constant plus an exponentially decreasing function (Eq. \ref{alpha}).

    
    \item Background noise activity, $\lambda_0\ll1$: is introduced to avoid logarithms of zero in the calculations resulting in all minicolumns having a minimal activity. 
    
    \item Network size, $H=M$: in a network with $N=H\times M$ units organized in $H$ hypercolumns with $M=H$ minicolumns each, represents the number of minicolumns available to store each memory. Based on experience, it is often good to have $H=M$ and this was used in the current study. Notably, our network models are tiny compared to, for instance, a biologically configured cortical area network which would rather have $H = 10000+$ and $M = 100$.

    \item Degree of memory cue perturbation, $s$: is the number of hypercolumn swaps in the perturbed pattern used to cue the network during recall, representing the dissimilarity of the cue to the target memory.
    
    \item Overlap threshold, $\theta_{o}$: the overlap between the final pattern reached and the target pattern needed for a successful recall, represents the required match of the retrieved memory with the encoded one.

\end{itemize}

\subsection{Simulation Protocol}

We sequentially encode several patterns in the network, each one representing the memories of one year of life, by clamping the activation of each pattern and letting $w$ and $b$ evolve during a predefined learning time.

Recall consists in letting the network state evolve during a predefined recall time after presenting it with a perturbed version of a pattern - the target pattern with $s$ perturbed hypercolumns having its active minicolumn randomly swapped. First part of recall comprises a predefined clamping recall time, whereafter the clamp is removed and the network dynamics evolve during the remaining recall time.

Recall overlap is the overlap between the actual pattern, $\pi_{true}$, a binary vector which length corresponds to the total number of  units in the network, and the final state reached, $\pi_{final}$, a vector of the same length with activities of corresponding units:

\begin{equation}\label{eqm7}
o=\frac{\pi_{true} . \pi_{final}}{||\pi_{true}||||\pi_{final}||}
\end{equation}

For each pattern, the network is presented several times with a different perturbed version of that pattern and the overlaps are calculated.

A successful recall is yielded when the overlap exceeds a certain fixed threshold, $\theta{o}$.
The ratio of retrieval of a pattern quantifies recall intensity:

\begin{equation}\label{eqm8}
r=\frac{number\; of\; successful\; recalls}{number\; of\; recall\; attempts}
\end{equation}

\begin{figure}[!ht]
\begin{center}
  \includegraphics[width=0.98\linewidth]{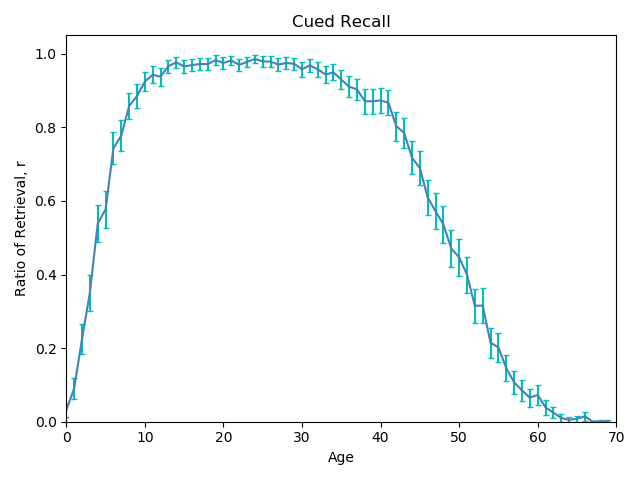}
  
 \end{center}
  \caption{Retrieval curve with initial configuration of parameters (Table \ref{t1}) and standard error of the mean}
  \label{f13}
  \end{figure}

This ratio $r$ is our measure of interest having direct interpretation as the portion of recalled memories for a given pattern. By plotting $r$ for each training pattern, corresponding to each age, we obtain a lifespan retrieval curve with the reminiscence bump (Figure  \ref{f13}). 

To obtain a smooth curve we plot the mean of several generated networks and several recall attempts per network, as reported in Table \ref{t1}, yielding a total of $100\times100=10000$ recall attempts to obtain the ratio of retrieval for a given year and $70\times10000=700000$ attempts to generate one single retrieval curve.

To investigate the effect of and sensitivity to a certain parameter on the reminiscence bump characteristics, a greedy approach is followed - all parameters are kept constant in the simulation except the one that is being investigated.

The initial configuration of simulation parameters is defined in Table \ref{t1}:

\begin{table}[!ht]
\centering
\caption{Parameters for the initial configuration of simulations}
~\\
\label{t1}
 \renewcommand{\arraystretch}{1.1}
\begin{tabular}{|l|c|}
   
\hline
\textbf{Parameter} & \textbf{Value} \\ 
\hline
Network size, $H\times M$  & $12 \times 12$ \\

Number of presented patterns & $70$ \\


Synaptic learning rate at birth, $\alpha_0$ & $0.3$\\

Time constant of the age-dependent plasticity decay, $\tau_s$ & $10$\\

Background activity level, $\lambda_0$ & $0.01$\\

Euler step, $dt$  & $0.01$\\

Learning time & $1$\\

Clamping recall time & $0.1$\\

Recall time & $2$\\

Degree of memory cue perturbation, $s$ & $6$\\

Number of generated networks & $100$\\

Number of perturbed patterns per age in each network & $100$\\

Overlap threshold, $\theta_{o}$ & $11/12$\\

Constant plasticity, $\alpha_{baseline}$ & $-$\\

\hline
\end{tabular}
\end{table}

\section{Results}

\subsection{Reminiscence Bump Replication}

BCPNN model was used to study the reminiscence bump feature of the memory retrieval curve. To this end, the network was simulated according to protocol described in section III. Figure 3 illustrates the memory retrieval curve generated with the model using the parameters listed in Table \ref{t1}. 

It exhibits a reminiscence bump coherent with the psychological studies and ideal shape and size for the subsequent simulations starting point, although it should be pointed out that memory performance close to 100\% is not reached in these studies.  

\subsection{Systematic Investigation on the Parametrization of the Reminiscence Bump}

\begin{figure*}
\begin{subfigure}{.5\linewidth}
  \centering
  \includegraphics[width=.95\linewidth]{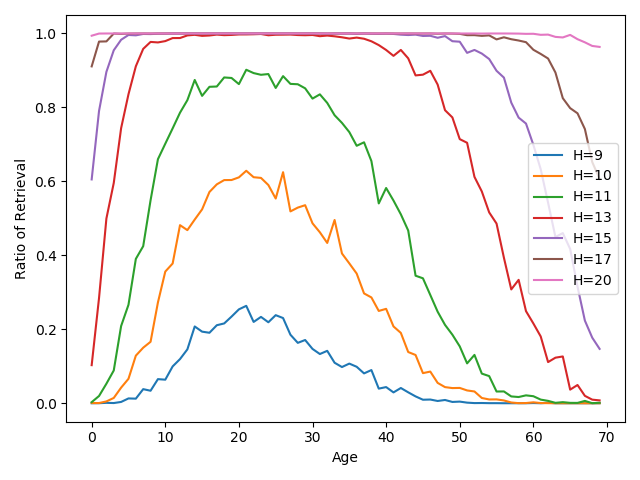}  
  \caption{Network size}
  \label{f14}
\end{subfigure}
\begin{subfigure}{.5\linewidth}
  \centering
  \includegraphics[width=.95\linewidth]{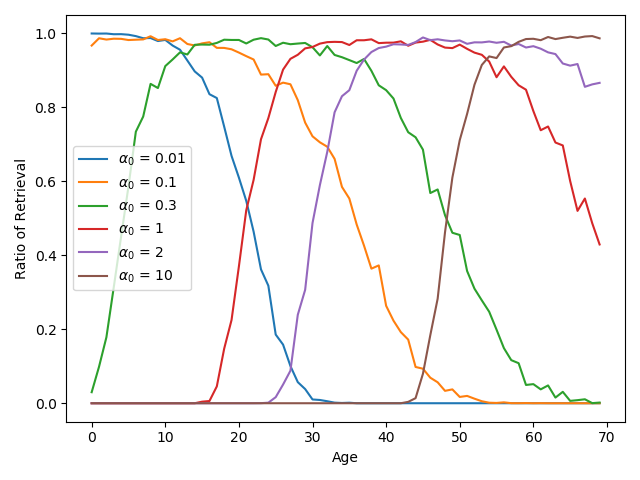}  
  \caption{Synaptic learning rate at birth}
  \label{f16}
\end{subfigure}

\begin{subfigure}{.5\textwidth}
  \centering
  \includegraphics[width=.95\linewidth]{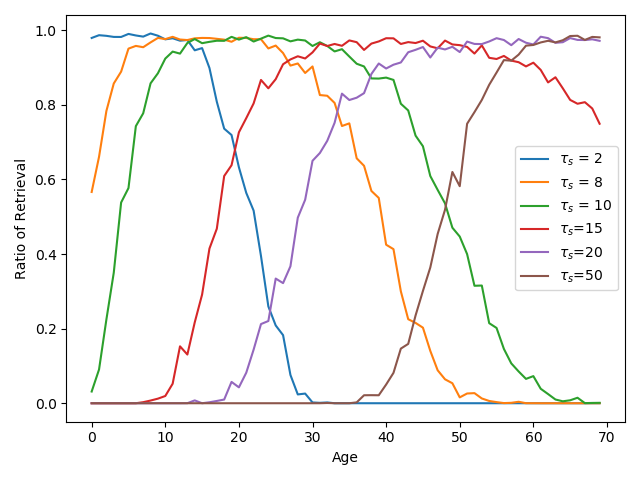}  
  \caption{Time constant of the age-dependent plasticity decay}
  \label{f18}
\end{subfigure}
\begin{subfigure}{.5\linewidth}
  \centering
  \includegraphics[width=.95\linewidth]{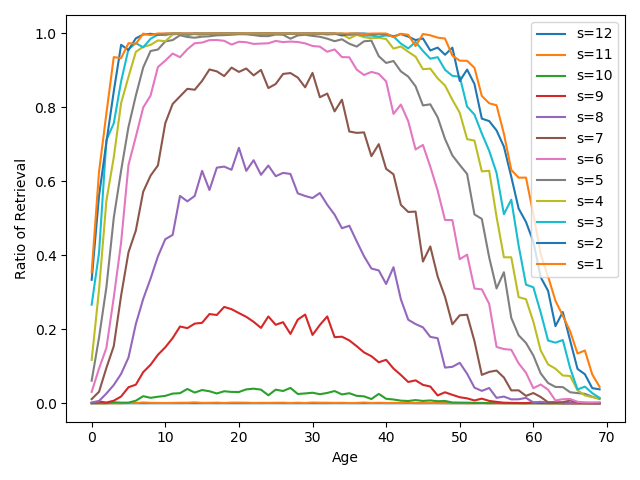}  
  \caption{Degree of memory cue perturbation}
  \label{f22}
\end{subfigure}
\caption{The memory retrieval performance over the network’s lifetime depending on different parameters}
\label{f4}
\end{figure*}

We then examined the effect of selected parameters on the characteristics of the reminiscence bump. These can be observed in Figure \ref{f4}.

\subsubsection{Network size, $H\times M$}

The larger the network size, the higher the ratio of retrieval (Figure \ref{f14}). This is due to the lower crosstalk between the stored memory patterns.

\subsubsection{Synaptic learning rate at birth, $\alpha_0$}

A shift of the bump is mediated by the synaptic learning rate for $age=0$ (Figure \ref{f16}). If it is low, the plasticity, the ability of the model to change its connections in order to learn new patterns, is very low at the older ages of the network, resulting in recalling only memories from early years. If it is too high, only memories from the recent years are recalled since there is a high level of plasticity despite the simulated decay (Eq. \ref{alpha}), which makes the network “overwrite” early memories.

\subsubsection{Time constant of the age-dependent plasticity decay, $\tau_s$}

The higher the decay time constant of plasticity is, the more the bump shifts to a later age (Figure \ref{f18}). It has the same effect as varying the synaptic learning rate at the simulated birth (compare with Figure \ref{f16}).

\subsubsection{Degree of memory cue perturbation, $s$}

To test the network's pattern completion capabilities we introduced binary swaps to the cue relative to the original training pattern. Unsurprinsingly, the stronger the perturbation the lower retrieval rate was reported. This is because highly perturbed patterns lead the network state to another attractor that does not correspond to the target one (Figure \ref{f22}).

\subsubsection{Overlap threshold, $\theta_{o}$}

The retrieval was robust and the detection was insensitive to the overlap threshold once it exceeded $5/12$.

\subsubsection{Background activity level, $\lambda_0$
}

It was observed that the lower the background activity level, the higher the recall. This effect was expected since this activity introduces noise.

\subsection{Modelling the Recency Phenomenom}

In an effort to model the recency phenomenom we set the synaptic learning rate decay to stop at a certain age which yielded the desired recency tail in the retrieval curve. 

With regard to model equations, we parameterized this phenomenon by decomposing the evolution of the learning rate as a constant plus an exponentially decreasing function. 

\begin{figure}[!ht]
\begin{center}
  \includegraphics[width=0.98\linewidth]{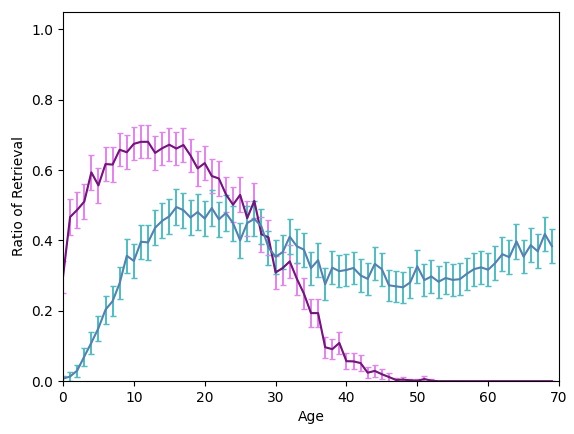}
 \end{center}
\caption{Recency with the standard error of the mean (blue) and same experiment without the constant baseline plasticity (purple)
}
\label{f26}
\end{figure}

The values of the parameters resulting in the recency graph (Figure \ref{f26}) are the values of Table \ref{t1} with the following changes:

\begin{itemize}
    \item synaptic learning rate at birth, $\alpha_0 = 0.25$
    \item time constant of the age-dependent plasticity decay, \newline $\tau_s = 8$
    \item constant plasticity, $\alpha_{baseline} = 0.015$
     \item degree of memory cue perturbation, $s = 8$
\end{itemize}

These changes were made to better fit the graphs from experimental studies with humans \cite{rubin1998things}. Childhood amnesia, the inability of adults to recall episodic memories from their early childhood, is also reproduced (Figure \ref{f26}). 

The same experiment without the constant baseline plasticity parameter was also performed. It can be observed that the introduction of the constant plasticity leads to a decrease in the recall of childhood memories and to the desired recency phenomenon effect (Figure \ref{f26}).

\begin{figure*}[ht]
\begin{subfigure}{.5\textwidth}
  \centering
  \includegraphics[width=.95\linewidth]{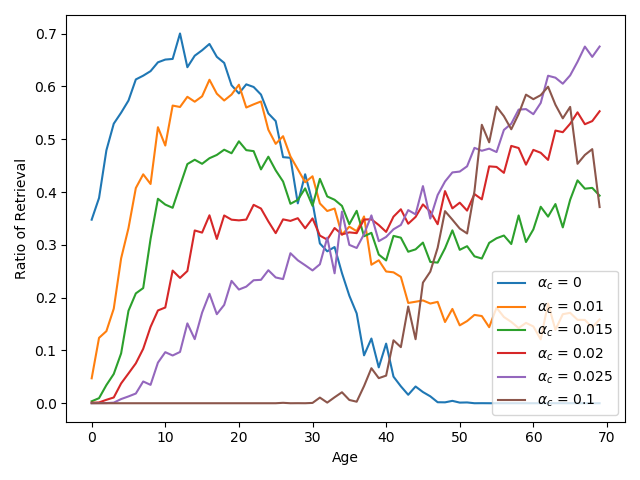}  
  \caption[width=.9\textwidth]{Constant baseline plasticity}
  \label{fr1}
\end{subfigure}
\begin{subfigure}{.5\textwidth}
  \centering
  \includegraphics[width=.95\linewidth]{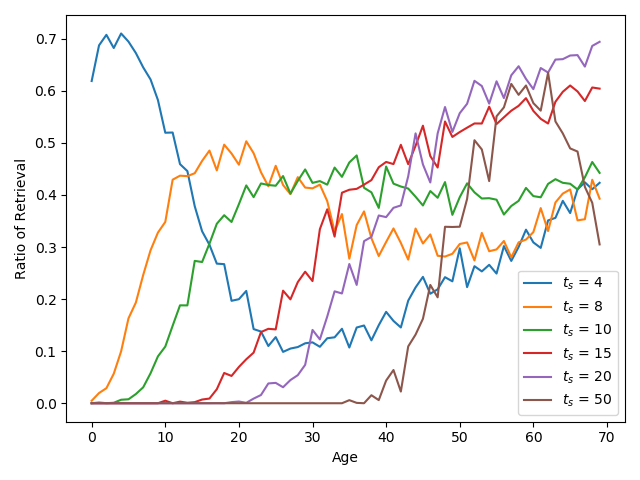}  
  \caption[width=.9\textwidth]{Time constant of the age-dependent plasticity decay}
  \label{fr2}
\end{subfigure}
\caption{The memory retrieval performance over the network’s lifetime depending on different parameters}
\label{fa}
\end{figure*}

We performed simulations in order to investigate the effects of varying the most relevant parameters in this setting, the constant plasticity and the time-constant of the age-dependent plasticity decay. These are depicted in Figure \ref{fa}.

\subsubsection{Constant baseline plasticity, $\alpha_{baseline}$  
}

The constant baseline plasticity is a fixed component of plasticity level (Eq. \ref{alpha}) that sets the lower limit for the decay of plasticity throughout lifetime. Thus, increasing constant plasticity results in a bump more shifted to the right and a higher recency effect. If the fixed plasticity is very high a bump forms in older ages due to the high plasticity at that age that prevails over the decreasing plasticity function.

\subsubsection{Time constant of the age-dependent plasticity decay, $\tau_s$}

The time constant of the age-dependent plasticity decay mediates the time it takes for the initial plasticity level to decrease (Eq. \ref{alpha}). It can be observed that increasing this parameter has a similar effect of increasing the constant plasticity (compare Figure \ref{fr1} with Figure \ref{fr2}) although the shape of the curve is slightly different.

\section{Discussion}

\subsection{Summary of Findings}

The parameters that showed the most substantial effect in the bump characteristics were the synaptic learning rate at birth and time constant of the age-dependent plasticity decay because these set the position of the bump, namely the peak age at which the retrieval curve has higher magnitude. The proposed constant component of the plasticity value throughout lifetime enables to model the recency phenomenon and also has a substantial effect on the shape of the lifespan retrieval curve. By tuning this parameter a curve with a recency tail is achieved. The other parameters have a lower relevance since they mainly only influence the magnitude of the retrieval curve.

\subsection{Interpretation of the Results}

Our modelling is consistent with the neurobiological hypothesis that the mechanisms underlying the reminiscence bump are:

\begin{itemize}
    \item the decrease of brain plasticity with aging due to dropping levels of dopamine receptors
    \item the pruning of synapses with aging
\end{itemize}

These are represented in the model by the exponentially decaying synaptic learning rate throughout time.

Dopamine D1 activation influences synaptic plasticity \cite{otani2003dopaminergic}. It can provoke neuronal excitation or inhibition, resulting in synaptic potentiation or depression, an increase or decrease in the efficacy of the synapses, or “connections” between neurons. It is known that dopamine D1 receptor density decreases with aging \cite{wang1998age}. 

By tuning the most important parameters - the synaptic learning rate at birth, time constant of the age-dependent plasticity decay and constant baseline plasticity - we could replicate quite precisely the lifespan retrieval bump including childhood amnesia, the reminiscence bump and the recency phenomenon at the ages, amplitudes and proportions consistent with data from experimental studies with humans \cite{rubin1998things}. 

There was no need for a cascade of systems or different encoding and forgetting functions such as in the attempt to replicate the reminiscence bump with the Memory Chain Model \cite{6} and the curve is similar to the curve generated by the AM-ART model \cite{7}.

The parametrization of the curve with the recency phenomenmon is still compatible with the neurobiological hypothesis of decreasing dopamine receptors with aging. Its parametrization using the constant plasticity parameter suggests a biologically motivated assumption that the dopamine decay throughout lifetime has a lower limit. 

\subsection{Limitations}

Our modelling approach considered solely long term memory, thus not considering the interactions between different brain areas as in the Memory Chain Model \cite{6}. In our model, these interactions could be represented by connecting several networks representing the different areas that deal with memory and making use of synaptic adaptation. This approach did not seem necessary for our modelling purposes but is a step towards a more complete model.

Furthermore, if we used a measure of how strongly encoded a pattern is, such as a sensitivity index, we could replicate experimental forgetting curves \cite{rubin1996one}, i.e. how strongly a pattern is encoded along time for each age.
By doing this, the values obtained would have a more relevant meaning and the analysis could be quantitatively more realistic.

The values used in our parametrization are thus less relevant. The qualitative relations are expected however to be interpretable. Thus, this approach allowed us to understand the origin of translation as well as decrease and increase of the bump magnitude, but the precise values play little role.

\section{Conclusions and Future Work}

In this work, we modelled the human lifespan retrieval curve with an incremental firing-rate attractor neural network model featuring a Bayesian-Hebbian learning rule \cite{39}. The effect of several parameters of the model were analyzed in a systematic way with the objective to study the mechanisms that modulate bump characteristics - age and magnitude - in this firing-rate attractor neural network model with BCPNN plasticity \cite{8}. 
The parameters that showed the most significant effect on the bump characteristics were the synaptic learning rate at birth and time constant of the age-dependent plasticity decay that set the position of the bump. 
We also proposed the introduction of a constant component of plasticity value to model the recency phenomenon. Such component also demonstrated a significant impact on the position of the bump and shape of the lifespan retrieval curve. 
The other parameters mainly influence the magnitude of the retrieval curve.

Despite the model’s simplicity and high level of abstraction it demonstrated considerable potential to simulate the human lifespan retrieval curve phenomena and provided insights into several mechanisms underlying reminiscence bump characteristics and even recency and childhood amnesia. It remains to study the effect of varying all the parameters in the final model including the recency phenomenon. Furthermore, a free recall setup of this model could show to what extent the effects of varying the parameters are conserved.
Future studies might further consider a more complete, detailed and spiking multi-network model featuring several different plasticity time constants spanning from seconds to minutes to years \cite{fiebig2014memory}. This would allow a better understanding of the interaction and coordination of the different memory systems in the healthy as well as dysfunctional brain. For a more detailed description of the simulation settings, additional experimental results and background in the lifespan retrieval curve refer to \cite{pereira2020attractor}.

\section*{Acknowledgements}

This work was partially supported by Fundação para a Ciência e a Tecnologia (FCT), through Portuguese national funds Ref. UIDB/50021/2020.


\end{document}